\documentclass[aps,pra,twocolumn,superscriptaddress,groupedaddress,showpacs,showkeys,floatfix]{revtex4}

\usepackage[dvips]{graphicx}
\usepackage{bm} 

 \makeatletter
 \newcounter{subeqncnt}
 \def\thesubeqncnt{\alph{subeqncnt}}%
  \def\subequations{\begingroup%
     \stepcounter{equation}\edef\@tempa{\theequation}%
     \let\c@equation\c@subeqncnt\c@subeqncnt\z@
     \edef\theequation{\@tempa\noexpand\thesubeqncnt}}
 
 \makeatother

\begin{document}
\title{Controlling phase separation of binary Bose-Einstein condensates
via mixed-spin-channel Feshbach resonance}
\author{Satoshi Tojo}
\affiliation{Department of Physics, Gakushuin University,
Tokyo 171-8588, Japan}
\author{Yoshihisa Taguchi}
\affiliation{Department of Physics, Gakushuin University,
Tokyo 171-8588, Japan}
\author{Yuta Masuyama}
\affiliation{Department of Physics, Gakushuin University,
Tokyo 171-8588, Japan}
\author{Taro Hayashi}
\affiliation{Department of Physics, Gakushuin University,
Tokyo 171-8588, Japan}
\author{Hiroki Saito}
\affiliation{Department of Engineering Science, University
of Electro-Communications, Tokyo 182-8585, Japan}
\author{Takuya Hirano}
\affiliation{Department of Physics, Gakushuin University,
Tokyo 171-8588, Japan}
\date{\today}

\begin{abstract}

We investigate
controlled phase separation of a binary Bose--Einstein condensate (BEC)
in the proximity of mixed-spin-channel Feshbach resonance
in the $|F=1,m_F=+1\rangle$ and $|F=2,m_F=-1\rangle$ states
of $^{87}$Rb at a magnetic field of 9.10 G.
Phase separation occurs on the lower magnetic-field side of 
the Feshbach resonance
while the two components overlap on the higher magnetic-field side.
The Feshbach resonance curve of the scattering length is obtained 
from the shape of the atomic cloud
by comparison with the numerical analysis of 
coupled Gross--Pitaevskii equations.

\end{abstract}

%
%
%

\maketitle

\section{Introduction}

Ultracold atomic gases 
provide an attractive testing ground for studying 
dynamics of multicomponent quantum fluids.
It has been shown that 
dual-species quantum gases \cite{Modugno, Ospelkaus, Papp},
two-component Bose--Einstein condensates (BECs)
comprised of two different hyperfine states \cite{Hall, Anderson}, 
and spinor BECs with different Zeeman sublevels \cite{StengerN,Miesner,StamperKurn}
exhibit a rich variety of dynamics.
The controllability of the intra and intercomponent
interactions via Feshbach resonance \cite{Inouye} 
creates numerous possibilities and enriches the physics of
multicomponent quantum fluids.

Miscibility between different components is crucially important to the
dynamics of multicomponent systems.
Phase separation in immiscible two-component BECs
has been studied in Ref.~\cite{Ho}.
Immiscible two-component BECs have been predicted to have 
interface instabilities:
the Kelvin--Helmholtz instability in the presence of shear flow \cite{Takeuchi}
and the Rayleigh--Taylor instability \cite{Sasaki}.
On the other hand,
the miscibility between different spin components
plays a key role in coherent spin dynamics,
such as the Josephson oscillation \cite{Chang, Kuwamoto},
spin echoes \cite{Yasunaga},
the Ramsey interferometer \cite{Anderson},
spin entanglement \cite{Teichmann,Gross},
and determination of the magnetic ground state of a spinor BEC~\cite{TojoPRA}.

Papp {\em et al}.~\cite{Papp} recently tuned the miscibility in binary
BECs of $^{85}$Rb--$^{87}$Rb 
controlling the intracomponent interaction of $^{85}$Rb.
The present system differs from that in Ref.~\cite{Papp} in that the
miscibility of different spin states of an identical species is
controlled by intercomponent Feshbach resonance.
For a binary mixture of two different internal states,
the populations of the components can be altered
at any stage in an experiment via spin manipulations.
Although the Feshbach resonance has been reported
on different internal states of an identical species
\cite{Kempen, Erhard, WideraFBR},
control of their miscibility has not been discussed.

The scattering length determines the properties of ultracold collisions
\cite{Weiner}.
Spectroscopic methods 
for determination of the scattering length
by observation of energy shift have been demonstrated
\cite{Roberts,Regal,Widera,Uetake}.
While these methods have advantages in accuracy, 
they are  applicable only 
to the states between which spectroscopic transition is available.
In addition,
the spectroscopic methods 
have a disadvantage for high density system, 
since an energy shift
caused by atomic density lowers the precision of the estimation.

In this paper,
we control phase separation via Feshbach resonance between internal spin states, 
$|F=1, m_F=1 \rangle \equiv |1\rangle$ 
and $|F=2, m_F=-1 \rangle \equiv |2\rangle$, of $^{87}$Rb.
The miscibility of these two components is found to 
depend sensitively on the strength of
the applied magnetic field near the Feshbach resonance.
The experimental results are compared with
numerical simulations of coupled Gross--Pitaevskii (GP) equations.
The excellent agreement between the experimental results and the numerical
simulations allows us to estimate the scattering length 
between the internal states.
The experiments and simulations for various values of magnetic field yield
the resonance curve of the scattering length near the
Feshbach resonance.
Thus observation of phase separation dynamics can be used as a new
method to estimate scattering lengths
of multicomponent BECs,
which does not rely on spectroscopic transition and can be used in the
high density regime.

This paper is structured as follows.
In Sec.\ref{sec2}, we introduce the mean-field formalism for a binary BEC.
In Sec.\ref{sec3}, our experimental apparatus and conditions are described.
In Sec.\ref{sec4}, experimental results are compared with numerical simulations.
Sec.\ref{sec5} is devoted to the conclusions.

\section{Two component condensates}\label{sec2}

The dynamics of a binary BEC with inelastic two-body losses
is described by coupled GP equations \cite{Hall, Anderson, TojoPRA},
\begin{subequations}
\begin{eqnarray}
\!\!\!\!\!\! i\hbar\frac{d \psi_1}{d t} \!\!\! &=& \!\!\! 
\left( -\frac{\hbar^2 \nabla^2}{2m}+V +
  \tilde{g}_{11}|\psi_1|^2 + \tilde{g}_{12}|\psi_2|^2 \right)
\psi_1, \\
\!\!\!\!\!\! i\hbar\frac{d \psi_2}{d t} \!\!\! &=& \!\!\! 
\left( -\frac{\hbar^2 \nabla^2}{2m}+V +
  \tilde{g}_{22}|\psi_2|^2 + \tilde{g}_{12}|\psi_1|^2 \right) \psi_2,
\end{eqnarray}
\label{cgp}
\end{subequations}
$\!\!\!\!$where
$\psi_{i}$ is the macroscopic wave function for the $|i\rangle$ state,
$m$ is the mass of $^{87}$Rb, and 
$V$ is the trap potential.
The interaction coefficient is given by
$\tilde{g}_{ij} = g_{ij} - i\hbar K_{ij}/2$ with
$g_{ij} = 4\pi \hbar^2 a_{ij}/m$,
where $a_{ij}$ is the scattering length and
$K_{ij}$ is the two-body inelastic collision rate
between the $|i\rangle$ and $|j\rangle$ states of condensates.
The scattering lengths of $^{87}$Rb have almost the same values:
$a_{22} =  95.00$ $a_{\rm B}$, $a_{11} = 100.4$ $a_{\rm B}$, and 
$a_{12} = 97.66$ $a_{\rm B} \equiv a_{\rm{bg}}$,
where $a_{\rm B}$ is the Bohr radius \cite{Hall}.
The scattering lengths determine the miscibility of the binary BEC. 
The phase separation condition in a uniform system is given by
$a_{12}^2 > a_{11} a_{22}$.
The atom density in two component condensates decreases by the two-body
inelastic collisions as
\begin{equation}
\frac{\partial n_{m}}{\partial t} = 
-K_{m m} n_{m}^2 
-K_{m m^{\prime}} n_{m} n_{m^{\prime}}.
\label{eqdensity}
\end{equation}
The two-body inelastic collision rates
in the $F=2$ manifold were measured 
in Ref. \cite{TojoPRA}, giving
$K_{22} = 1.04 \times 10^{-13}$cm$^3$/s.
The two-body inelastic loss in the $F=1$ manifold is negligible and we
assume $K_{11}=0$.

Near the magnetic Feshbach resonance, 
the intercomponent interaction $\tilde{g}_{12}$ is changed.
The interspecies scattering length in 
a complex form can be expressed as 
a Lorentzian function \cite{Hutson}.
Since the imaginary part of the scattering length is incorporated in $K_{12}$, 
we use the effective scattering length 
between the $|1\rangle$ and $|2\rangle$ states defined by
\begin{equation}
a_{12}^{\rm{eff}} \equiv a_{\rm bg} + \Delta a(B)
= a_{\rm bg} \left[1- \frac{\Delta B (B-B_0)}{(B-B_0)^2 + (\gamma_{\rm B}/2)^2 }\right],
\label{eqscat}
\end{equation}
where the parameters $B_0$, $\Delta B$, and $\gamma_{\rm B}$ are determined later.

\section{Experiment}\label{sec3}

Our experimental apparatus and procedure used to create $^{87}$Rb condensates
are the same as those described in Refs.~\cite{Kuwamoto, TojoPRA}
except for irradiation of a two-photon $\pi$/2 pulse 
between the hyperfine states.
A BEC containing $10^6$ atoms in the $|2,2\rangle$ state is
created by evaporative
cooling with frequency sweeping of an rf field for 18 s in a magnetic trap.
The BEC is loaded into a crossed far-off resonant optical trap (FORT) 
at a wavelength of 850 nm and $3\times10^{5}$ atoms remain in the FORT.
The potential depth of the crossed FORT is estimated to be about 1 $\mu$K
and the radial (axial) trap frequency is measured to be 141 Hz (21 Hz) 
using the parametric resonance.
After sudden inversion of the quantization axis,
the $|2,-2\rangle$ state is transferred to 
the $|2,-1\rangle \equiv |2\rangle$ state
by inducing the Landau--Zenner transition using rf irradiation
with an external magnetic field of 20.5 G. 
Half the atoms in the 
$|2\rangle$ state is then
transferred to the $|1,1 \rangle \equiv |1\rangle$ state
by irradiation of a $\pi$/2 pulse of
rf and microwave field for 5 ms.
After time evolution for $t_{\rm ev}$
in a precisely controlled magnetic field $B_{\rm ev}$,
the crossed FORT is turned off.
The magnetic field is kept on 
for the first 5 ms of the time-of-flight (TOF)
to maintain the 
scattering length during expansion of the atomic cloud.
After applying a Stern--Gerlach pulse,
absorption images are obtained.
In order to take the image of each component, we apply
a sequence of two short 
pulses after a TOF times of 15 ms for $F=2$ 
and 18 ms for $F=1$ with repumping to $F=2$.
The fluctuation in the number of atoms is estimated to be 10\%.
The relative population fluctuates
between 0.45 and 0.55 in each run of the experiment
\cite{TojoAPB,TojoPRA}.

%
%
\begin{figure}
\begin{center}
\includegraphics[width=6.5cm]{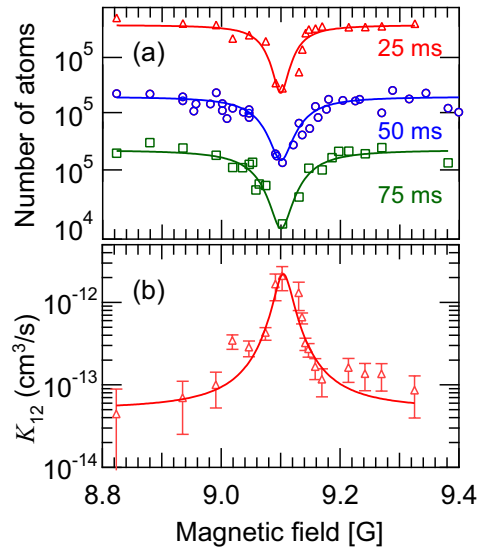}
\caption{(Color online) 
(a) Magnetic-field dependence of 
the atom number for evolution times $t_{\rm ev}$ of 25, 50, and 75 ms.
(b) Inelastic collision rate estimated by Eq.(\ref{eqdensity}) at $t_{\rm ev}=$ 25 ms.
Solid lines are Lorentzian functions fit to the data with center at 9.10 G.
}
\label{atoms}
\end{center}
\end{figure}

The magnetic-field strength of the Feshbach resonance
can be determined by measuring the atomic losses.
In our experimental conditions, 
atomic losses by one- and three-body inelastic collisions
are negligible compared with that by two-body 
inelastic collisions \cite{TojoPRA}.
The total number of atoms in the
$|2\rangle$ state decreases rapidly with increasing trap time $t_{\rm ev}$ 
because of hyperfine-changing inelastic collisions such as
$|2,-1\rangle$, $|2,-1\rangle \rightarrow |1,m_F\rangle$, 
$|1$ or $2,m_{F}^{'}\rangle$ and
$|2,-1\rangle$, $|1,-1\rangle \rightarrow |1,m_F\rangle$,
$|1,m_{F}^{'}\rangle$.
For a magnetic field far from the Feshbach resonance,
the two-body inelastic
collision rate between the $|2\rangle$ and $|1\rangle$ states is
estimated to be $K_{12} = 0.5 \times 10^{-13}$cm$^3$/s, 
which is obtained by comparing the atom loss in our
experiment and a solution of the GP equation (1).
Both the elastic scattering length and the inelastic collision loss rate
are altered in the vicinity of the Feshbach resonance \cite{Inouye}.
Figure \ref{atoms}(a) shows the total number of atoms 
after the TOF for evolution times $t_{\rm ev}$ of 25, 50, and 75 ms.
%
%
\begin{figure*}
\begin{center}
\includegraphics[width=14.0cm]{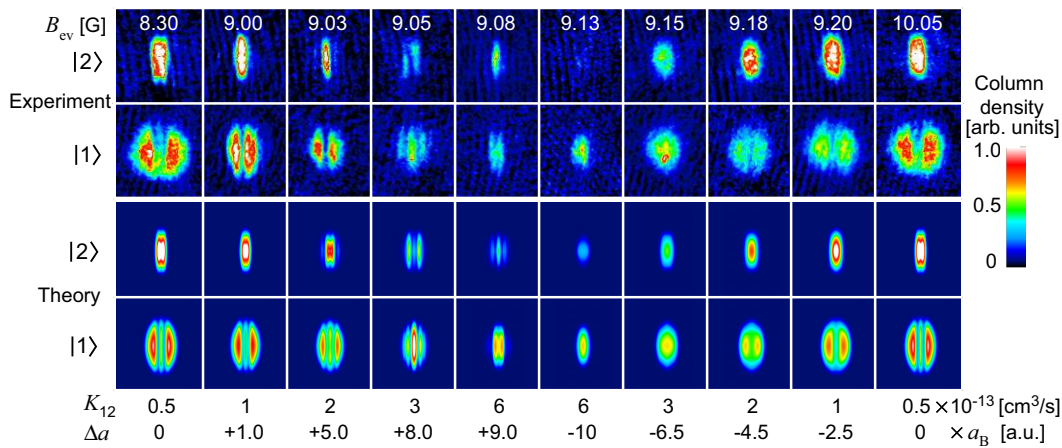}
\caption{(Color online) 
Column densities of the $|2\rangle$ and $|1\rangle$ states
obtained by the experiment (upper) and
the numerical simulation (lower)
as a function of magnetic field $B_{\rm ev}$. 
The field of the view is 316 (vertical) $\times$ 290 $\mu$m (horizontal).
The two-body inelastic collision rates $K_{12}$ ($10^{-13}$cm$^3$/s) 
and the change in the scattering lengths $\Delta a$ 
used in the numerical simulation are indicated at the bottom.
}
\label{density}
\end{center}
\end{figure*}
For $t_{\rm ev}$ = 25 ms, 
density profiles were almost the same at both lower and higher magnetic
field near the Feshbach resonance.
We estimated $K_{12}$ for $t_{\rm ev}$ = 25 ms by solving
Eq.~(\ref{eqdensity}) numerically with the single-mode approximation.
The uncertainty arises from the fluctuation in
the initial number of atoms.
The data are fitted by a Lorentzian function
as shown in Fig.~\ref{atoms}(b),
which gives Feshbach resonance field $B_0$
to be 9.100 G and $\gamma_{\rm B}$ to be 30 mG. 
The fluctuation in the magnetic field $B_{\rm ev}$ is estimated to be less than 5 mG 
by observing the magnetic dipole transitions,
and the residual gradient magnetic field is estimated
to be 30 mG/cm \cite{Kuwamoto}.
The magnetic-field strength is calibrated by
Rabi spectra between clock states with the microwave irradiation.
The uncertainty in this calibration is 5 mG.
The resonant magnetic field obtained in our experiment, 9.100(5) G,
agrees with the theoretical prediction \cite{Kaufman}
within the experimental uncertainty.

\section{Results and discussion}\label{sec4}

The column densities of the binary BEC around 
the Feshbach resonance for $t_{\rm ev}$= 75 ms are shown in Fig.~\ref{density}.
At a magnetic field far from the Feshbach resonance
($B_{\rm ev} =$ 8.30 and 10.05 G),
the two components exhibit phase separation \cite{Hall, Anderson}.
The domain structure of phase separation depends
not only on the scattering lengths
but also on the number of atoms and the relative populations.
The scattering lengths and inelastic collision rates for
$B_{\rm ev}=$ 8.30 G and those for $B_{\rm ev}=$ 10.05 G are almost the same,
since their density patterns are similar.
The behaviors of the binary BEC in the vicinity of 
the Feshbach resonance at 9.10 G change dramatically in Fig. \ref{density}
due to the change in the scattering length and the inelastic collision rate.
The domain structures on the lower magnetic-field side
near the Feshbach resonance ($B_{\rm ev} =$ 9.05 and 9.08 G)
are quite different from that at $B_{\rm ev} =$ 8.30 G.
On the other hand, 
the domain structure disappears at higher magnetic fields around
the Feshbach resonance.
The behavior at 9.15 G is quite different from that at 9.05 G
even though the numbers of atoms are almost the same.
These results indicate that the scattering length $a_{12}$ significantly
changes at $B_{\rm ev} \simeq$ 9.1 G.
The domain structures depicted in Fig.~\ref{density} are
reproducible within the experimental fluctuations in the initial number
of atoms and the relative populations.

%
%
\begin{figure}
\begin{center}
\includegraphics[width=8.0cm]{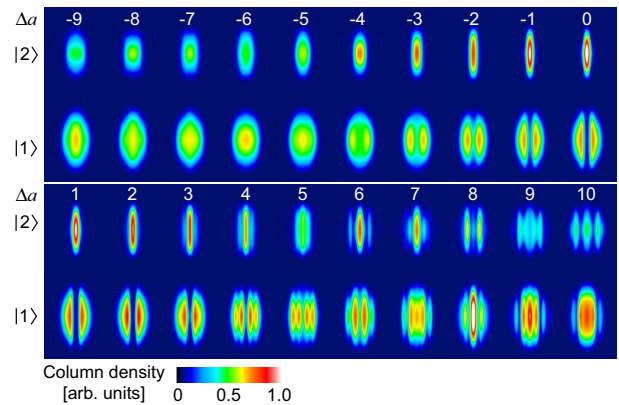}
\caption{(Color online) Numerically obtained column density distributions
for various scattering lengths at an evolution time of 75 ms
with $K_{12} = 3\times 10^{-13}$cm$^3$/s.
The relative scattering length $\Delta a$
in unit of $a_{\rm B}$ is indicated in the top row.
}
\label{sct}
\end{center}
\end{figure}

\begin{figure}
\begin{center}
\includegraphics[width=6.5cm]{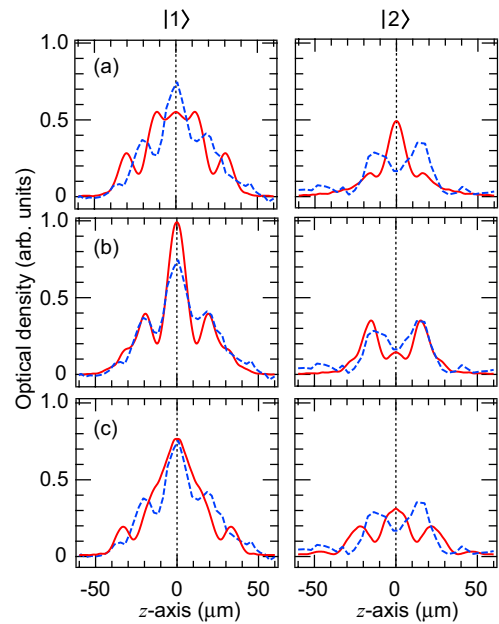}
\caption{
(Color online) Comparison between the experimentally observed density distributions at 
$B_{\rm ev} = 9.05$ G (dashed lines) and numerical ones (solid lines) obtained
for (a) $\Delta a = 7.5$ $a_{\rm B}$, (b) $8.0$ $a_{\rm B}$, and (c)
 $8.5$ $a_{\rm B}$. 
}
\label{cmp30}
\end{center}
\end{figure}

\begin{figure}
\begin{center}
\includegraphics[width=6.5cm]{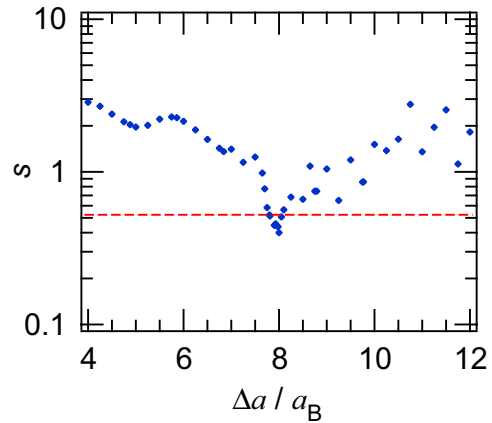}
\caption{
(Color online) 
Root-mean-square deviation
between experiments and calculations at
9.05 G with $K_{12} = 3 \times 10^{-13}$cm$^3$/s.
The dashed line shows 1.3 $s_{\rm min}$,
which defined as a threshold for determination of the error bar
of the scattering length.
}
\label{chisq}
\end{center}
\end{figure}

We numerically solve coupled GP equations (1) for various
values of $\Delta a$ and $K_{12}$, 
and obtain the column densities of atomic clouds after the TOF
(lower panels of Fig.~\ref{density}).
The GP equations were solved by the Crank-Nicolson method.
The numerically obtained column densities are smoothed to 
account for the spatial resolution (3.2 $\mu$m) of the CCD camera.
The values of $\Delta a$ and $K_{12}$ are determined as follows.
First, we assume $\Delta a = 0$ and calculate the time evolution of the
number of atoms for various $K_{12}$ using coupled GP equations (1).
We then compare the results with the experimental ones 
and estimate the value of $K_{12}$.
We confirmed that the difference between the number of atoms for 
$\Delta a = 0$ and that for $\Delta a \neq 0$ differ by less than 15 $\%$,
which is comparable to the fluctuation in the initial number of atoms
and does not affect the estimation of $K_{12}$.
We next calculate the column densities for various values of $\Delta a$,
compare them with the experimental results,
and estimate the value of $\Delta a$.
For example, 
Fig.~\ref{sct} shows a catalog of the column densities for
$K_{12} = 3 \times 10^{-13}$cm$^3$/s, 
which corresponds to
$B_{\rm ev} =$ 9.05 G and 9.15 G in Fig.~\ref{density}.
Comparing the experimental column densities in Fig.~\ref{density} 
with numerical ones in Fig.~\ref{sct},
we can estimate scattering lengths.
We assume that the most probable scattering length minimizes
the root-mean-square deviation
$s = \sqrt{\sum_z^{n}[\alpha_{\rm exp}(z)-\alpha_{\rm cal}(z)]^2/n}$,
where $\alpha_{\rm exp}(z)$ and $\alpha_{\rm cal}(z)$ are
optical densities in an experiment and a calculation at $z$,
respectively.
Figure \ref{cmp30} shows integrated optical densities
obtained by the experiment at 9.05 G and 
numerical calculation at 
$\Delta a =$ 7.5, 8.0, and 8.5 $a_{\rm B}$.
We find that the phase separation dynamics is sensitive to 
the change in the scattering length by 0.50 $a_{\rm B}$.
The difference between experiments and calculations 
are expressed by $s$ values as shown in Fig.~\ref{chisq}.
The value of $\Delta a$ that minimizes 
the value of root-mean-square deviation $s_{\rm min}$
is $\Delta a =$ 8.0 $a_{\rm B}$ for $B_{\rm ev}=$ 9.05 G.
In the range of $\Delta a > 8.0$ $a_{\rm B}$,
fluctuation of $s$ is expected to be larger than 
that in $\Delta a < 8.0$ $a_{\rm B}$.
This is because the density distribution of each component forms complex 
structures at $\Delta a > 8.0$ $a_{\rm B}$
while it shows simple domain structure in $\Delta a < 8.0$ $a_{\rm B}$.
The phase-separation dynamics can thus be used as a probe
for estimating the scattering length in a binary BEC.

The density patterns are sensitive to 
the differences in the scattering length for positive $\Delta a$.
The accuracy of the estimated scattering lengths
is therefore typically $\pm 1$ $a_{\rm B}$ 
for $\Delta a \gtrsim 3$ $a_{\rm B}$.
However,
in the close vicinity of the Feshbach resonance, 
the domain structure becomes moderate
since the number of atoms decreases considerably,
which makes accurate estimation of $\Delta a$ difficult.
For $-7$ $a_{\rm B} \lesssim \Delta a < 0$ in Fig.~\ref{sct},
inhomogeneous density distribution is formed,
even though the phase separation condition is not satisfied;
we can estimate $\Delta a =-6.5$ $a_{\rm B}$ at $s_{\rm min}$
for $B_{\rm ev} =$ 9.15 G.
This is because the number of $|2\rangle$ atoms decreases rapidly,
giving rise to a nonequilibrium density distribution.
Clear phase separation is necessary for precise estimation of 
the scattering length between $|1\rangle$ and $|2\rangle$.
In our experiment, 
$t_{\rm ev} <$ 25 ms is insufficient for observing the phase separation.
For $t_{\rm ev}>$ 100 ms,
the number of atoms in the $|2\rangle$ state becomes too small to
estimate $\Delta a$.
The evolution time of $t_{\rm ev} = $ 75 ms is the most suitable
to estimate the scattering length. 

%
%
%
%
%
%

For a uniform system, the Bogoliubov excitation spectrum has the form 
\cite{Pethick},
\begin{equation}
(\hbar \omega)^2 
\!\! = \! \varepsilon \!\! \left[ \varepsilon \! + \! g_{11} n_1 \! + \! g_{22}
n_2 \! \pm \!\! \sqrt{\!(g_{11} n_1 \! - \! g_{22} n_2)^2 \!\! + \! 4 n_1 n_2
g_{12}^2} \right],
\label{bogo}
\end{equation}
where $n_j = |\psi_j|^2$ is the atom density and $\varepsilon = \hbar^2
k^2 / (2m)$ with $k$ being the excitation wave number.
When $g_{12}^2 > g_{11} g_{22}$, $\omega$ is imaginary for 
$\varepsilon < [(g_{11} n_1 - g_{22} n_2)^2 + 4 n_1 n_2 g_{12}^2]^{1/2} - (g_{11}
n_1 + g_{22} n_2) $ 
and the system becomes
dynamically unstable against phase separation.
The exponential growth of the unstable mode is approximately given by
\begin{equation} \label{eq3}
\exp \left[ \int dt {\rm Im} \omega (t) \right],
\end{equation}
where $\omega(t)$ depends on time since $n_1$ and $n_2$ decrease with time.
The wavelength that maximizes Eq.~(\ref{eq3}) is most unstable,
which is, for example,
estimated to be 8.9 $\mu{\rm m}$ for 
$K_{12} = 1 \times 10^{-13}$cm$^3$/s and 
$\Delta a = 1.0$ $a_{\rm B}$ corresponding to 9.00 G,
and 4.9 $\mu{\rm m}$ for $K_{12} = 3 \times 10^{-13}$cm$^3$/s and 
$\Delta a = 8.0$ $a_{\rm B}$ to 9.05 G in Fig.~\ref{density}, respectively.
These estimations of the most unstable 
wavelengths are in good agreement
with the numerical solutions of Eq.~(1) before the TOF.
By the TOF expansion, the wavelengths
in the density pattern become a few times larger.

%
%
\begin{figure}
\begin{center}
\includegraphics[width=6.5cm]{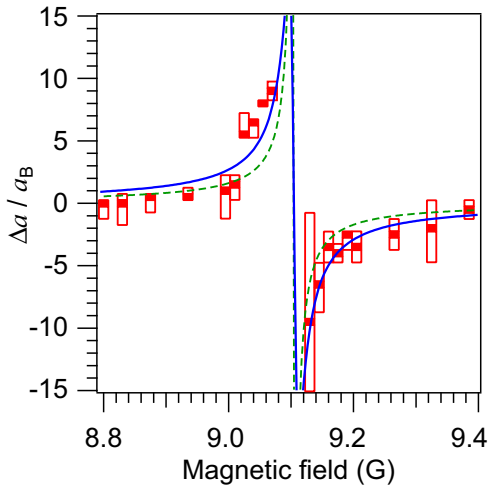}
\caption{
(Color online) Estimated value of the change in the scattering length
between the $|1\rangle$ and $|2\rangle$ states
for a magnetic field around the Feshbach resonance.
The scattering lengths at $s_{\rm min}$ are shown as filled
squares, and fitted by Eq.(\ref{eqscat}) (solid curve).
The open squares correspond to $s$
below the threshold value of 1.3 $s_{\rm min}$.
Theoretical prediction in Ref.\cite{Kempen} is indicated by the dashed curve.
}
\label{pltsct}
\end{center}
\end{figure}

Figure \ref{pltsct} depicts
the normalized scattering length $\Delta a$.
The theoretical prediction in Ref.~\cite{Kempen}
is adjusted to the resonance magnetic field of 9.10 G obtained in our
experiment (dashed curve).
The scattering lengths are obtained by the method in Fig.\ref{chisq}.
The filled squares correspond to $s_{\rm min}$
and the open squares show accuracy ranges below 
1.3 $s_{\rm min}$ of the threshold.
They are best fitted by Eq.(\ref{eqscat}) (solid line) 
with $B_0 = 9.104$ G 
($\gamma_{\rm B} = 13$ mG, $\Delta B = 3$ mG).
The value of $\Delta B$ is in good agreement with 
that of the theoretical prediction and other experiments
\cite{Kempen, Erhard, WideraFBR, Kaufman}.
The discrepancy in $\gamma_{\rm B}$ may be caused by
uncertainties near the resonance field.

The accuracy of $\Delta a$ in our method is 
comparable to that in the method using the Ramsey fringe
~\cite{WideraFBR}.
At $B_{\rm ev} > B_0$ with high density regime,
the accuracy of $\Delta a$ in our method
is improved at large $\Delta a$
because the phase separation dynamics is sensitive to $\Delta a$.
On the other hand, the accuracy in the Ramsey fringe method
at large $\Delta a$ becomes worse owing to 
density inhomogeneity caused by dramatic phase separation.
We note that the discrepancies in the scattering length
between experimental estimation and theoretical prediction 
may be caused by 
entangled spin states or molecular states near the Feshbach resonance
\cite{Teichmann, Gross}.
If these effects are taken into account,
the accuracy of the method may be improved.

\section{Conclusions}\label{sec5}

In conclusion, we observed the time evolution of binary $^{87}$Rb condensates
in the $|2,-1\rangle$ and $|1,1\rangle$ hyperfine states around the
Feshbach resonance at 9.10 G. 
In the vicinity of the Feshbach resonance,
the miscibility of the two components
is tuned to be both immiscible and miscible.
Phase separation occurs on the lower field side 
of the Feshbach resonance, while miscible behavior is observed on the 
higher field side.
We performed numerical simulations using coupled GP equations
and proposed a new method for determination of 
the scattering length.
We estimated the values of $\Delta a $ and $K_{12}$ by comparing the
experimental and numerical density distributions of the atomic cloud.
From systematic experiments and simulations,
we obtained the resonance curve of scattering length around the
Feshbach resonance.

Our method for determination of the scattering length
can be used for not only spectroscopic states
but also non-spectroscopic states,
and is powerful technique for high density regime
in both identical and different isotopes.
Miscibility control via mixed-spin-channel Feshbach resonance will open
up new possibilities for multicomponent quantum fluids,
such as controlled quantum phase transition between miscible and
immiscible phases
with precise tuned scattering lengths.
In addition,
this technique can be applied for a precise measurement of
a magnetic field below sub-milligauss range
in the case when the scattering length curve are well-known.

\begin{acknowledgments}

We would like to thank 
T. Kuwamoto, T. Tanabe, K. Hamazaki, M. Iwata, and E. Inoue
for their experimental assistance.
We also thank M. Tsubota for valuable discussions.
This work was supported by 
the Sumitomo Foundation,
Grants-in-Aid for Scientific Research
(No. 17071005, No. 18684024, No. 19740248, No. 20540388, and No. 22340116)
from the Ministry of Education, Culture, Sports, Science, and Technology
of Japan, 
and the Japan Society for the Promotion of
Science (JSPS) through its gFunding Program for World-Leading
Innovative R\&D on Science and Technology (FIRST Program)." 

\end{acknowledgments}


\begin{thebibliography}{99}
\bibitem{Modugno}
	G. Modugno, M. Modugno, F. Riboli, G. Roati, and M. Inguscio,
	Phys. Rev. Lett. \textbf{89}, 190404 (2002).
\bibitem{Papp}
	S. B. Papp, J. M. Pino, and C. E. Wieman,
	Phys. Rev. Lett. \textbf{101}, 040402 (2008).
\bibitem{Ospelkaus}
	S. Ospelkaus, C. Ospelkaus, L. Humbert, K. Sengstock, and K. Bongs,
	Phys. Rev. Lett. \textbf{97}, 120403 (2006).
\bibitem{Hall}
	D. S. Hall, M. R. Matthews, J. R. Ensher, C. E. Wieman, and
	E. A. Cornell,
	Phys. Rev. Lett. \textbf{81}, 1539 (1998);
	K. M. Mertes, J. W. Merrill, R. Carretero-Gonz\'alez, 
	D. J. Frantzeskakis, P. G. Kevrekidis, and D. S. Hall,
 	{\it ibid}. \textbf{99}, 190402 (2007).
\bibitem{Anderson}
	R. P. Anderson,
	C. Ticknor, A. I. Sidorov, and B. V. Hall,
	Phys. Rev. A \textbf{80}, 023603 (2009).
\bibitem{StengerN}
	J. Stenger,
	S. Inouye, D.M. Stamper-Kurn, H.-J. Miesner,
	A.P. Chikkatur, and W. Ketterle,
	Nature \textbf{396}, 345 (1998).
\bibitem{Miesner}
	H.-J. Miesner,
	D.M. Stamper-Kurn, J. Stenger, S. Inouye, A.P. Chikkatur, and
	W. Ketterle, 
	Phys. Rev. Lett. \textbf{82}, 2228 (1999).
\bibitem{StamperKurn}
	D.M. Stamper-Kurn,
	H.-J. Miesner, A.P. Chikkatur, S. Inouye, J. Stenger, and W. Ketterle, 
	Phys. Rev. Lett. \textbf{83}, 661 (1999).
\bibitem{Inouye}
	S. Inouye,
	M.R. Andrews, J. Stenger, H.-J. Miesner,
	D.M. Stamper-Kurn, and W. Ketterle,
	Nature \textbf{392}, 151 (1998).
\bibitem{Ho}
	T.L. Ho and V.B. Shenoy, Phys. Rev. Lett. \textbf{77}, 3276 (1996);
	B.D. Esry,
	C.H. Greene, J.P. Burke, Jr., and J.L. Bohn,
	{\it ibid}. \textbf{78}, 3594 (1997);
	E. Timmermans, {\it ibid}. \textbf{81}, 5718 (1998);
	H. Pu and N.P. Bigelow, {\it ibid}. \textbf{80}, 1130 (1998).
\bibitem{Takeuchi}
	H. Takeuchi,
	N. Suzuki, K. Kasamatsu, H. Saito, and M. Tsubota,
	Phys. Rev. B \textbf{81}, 094517 (2010).
\bibitem{Sasaki}
	K. Sasaki,
	N. Suzuki, D. Akamatsu, and H. Saito,
	Phys. Rev. A \textbf{80}, 063611 (2009).
\bibitem{Chang}
	M.-S. Chang,
	Q. Qin, W. Zhang, L. You, and M.S. Chapman,
	Nature Phys. \textbf{1}, 111 (2005).
\bibitem{Kuwamoto}
	T. Kuwamoto,
	K. Araki, T. Eno, and T. Hirano,
	Phys. Rev. A \textbf{69}, 063604 (2004).
\bibitem{Yasunaga}
	M. Yasunaga and M. Tsubota,
	Phys. Rev. Lett. \textbf{101}, 220401 (2008).
\bibitem{Teichmann}
	N. Teichmann and C. Weiss,
	Eur. Phys. Lett. \textbf{78}, 10009 (2007).
\bibitem{Gross}
	C. Gross, T. Zibold, E. Nicklas, J. Est\`eve, and M.K. Oberthaler,
	Nature \textbf{464}, 1165 (2010).
\bibitem{TojoPRA}
	S. Tojo, T. Hayashi, T. Tanabe, T. Hirano, 
	Y. Kawaguchi, H. Saito, and M. Ueda,
	Phys. Rev. A \textbf{80}, 042704 (2009).
\bibitem{Kempen}
	E.G.M. van Kempen,
	S.J.J.M.F. Kokkelmans, D.J. Heinzen, and B.J. Verhaar,
	Phys. Rev. Lett. \textbf{88}, 093201 (2002).
\bibitem{Erhard}
	M. Erhard,
	H. Schmaljohann, J. Kronj\"ager, K. Bongs, and K. Sengstock,
	Phys. Rev. A \textbf{69}, 032705 (2004).
\bibitem{WideraFBR}
	A. Widera,
	O. Mandel, M. Greiner, S. Kreim, T. W. H\"ansch, and I. Bloch,
	Phys. Rev. Lett. \textbf{92}, 160406 (2004).
\bibitem{Weiner}
	J. Weiner, V.S. Bagnato, S. Zilio, and P.S. Julienne,
	Rev. Mod. Phys. \textbf{71}, 1 (1999).
\bibitem{Roberts}
	J.L. Roberts,
	N.R. Claussen, J.P. Burke, Jr., C.H. Greene,
	E.A. Cornell, and C.E. Wieman,
	Phys. Rev. Lett. \textbf{81}, 5109 (1998);
	J.L. Roberts,
	N.R. Claussen, S.L. Cornish, and C.E. Wieman,
	{\it ibid}. \textbf{85}, 728 (2000);
	D.M. Harber, H.J. Lewandowski, J.M. McGuirk, and E.A. Cornell,
	Phys. Rev. A \textbf{66}, 053616 (2002).
\bibitem{Regal}
	C. A. Regal and D. S. Jin, 
	Phys. Rev. Lett. \textbf{90}, 230404 (2003).
\bibitem{Widera}
	A. Widera,
	F. Gerbier, S. F\"olling, T. Gericke, O. Mandel, and I. Bloch,
	New J. Phys. \textbf{8}, 152 (2006).
\bibitem{Uetake}
	S. Uetake,
	A. Yamaguchi, S. Kato, T. Fukuhara, S. Sugawa, K. Enomoto,
	Y. Takasu, and Y. Takahashi,
	{\it Proceeding of the 9th international Symbosium on
	Foundations of Quantum Mechanics in the light of new
	technology}, 12 (2008).
\bibitem{Hutson}
	T. K\"ohler, K. Goral, and P. S. Julienne,
	Rev. Mod. Phys. \textbf{78}, 1311 (2006);
	J. M. Hutson, New J. Phys. \textbf{9}, 152 (2007).
\bibitem{TojoAPB}
	S. Tojo,
	A. Tomiyama, M. Iwata, T. Kuwamoto, and T. Hirano,
	Appl. Phys. B \textbf{93}, 403 (2008).
\bibitem{Kaufman}
	A. M. Kaufman,
	R. P. Anderson, T. M. Hanna, E. Tiesinga, P. S. Julienne, and D. S. Hall,
	Phys. Rev. A \textbf{80}, 050701(R) (2009).
\bibitem{Pethick}
	C. J. Pethick and H. Smith,
	\textit{Bose-Einstein condensation in dilute gases},
	(Cambridge University Press, Cambridge, 2002) , Sect. 12.
\end{thebibliography}
%

\end{document}